\documentclass{llncs}
\usepackage{macros}

\hypersetup{%
pdftitle={Tactics for Reasoning modulo AC in Coq},%
pdfauthor={Thomas Braibant, Damien Pous},%
pdfkeywords={Coq proof assistant, decision procedure reflexive tactic, rewriting, modulo AC, associativity, commutativity},%
}

\title{Tactics for Reasoning modulo AC in Coq %
  \thanks{To appear \emph{in Proc. CPP}, \emph{LNCS}, Springer, 2011.}}%
\author{Thomas Braibant \and Damien Pous} %
\institute{LIG, UMR 5217, CNRS, INRIA, Grenoble} %

\begin{document}

\maketitle

\begin{abstract}
  We present a set of tools for rewriting modulo associativity and
  commutativity (AC) in Coq, solving a long-standing practical
  problem. We use two building blocks: first, an extensible reflexive
  decision procedure for equality modulo AC; second, an OCaml plug-in
  for pattern matching modulo AC.
  We handle associative only operations, neutral elements,
  uninterpreted function symbols, and user-defined equivalence
  relations.
  By relying on type-classes for the reification phase, we can infer
  these properties automatically, so that end-users do not need to
  specify which operation is A or AC, or which constant is a neutral
  element.
\end{abstract}

\section{Introduction}

\paragraph*{Motivations.}

Typical hand-written mathematical proofs deal with commutativity and
associativity of operations in a liberal way. Unfortunately, a proof
assistant requires a formal justification of all reasoning steps, so
that the user often needs to make boring term re-orderings before
applying a theorem or using a hypothesis.
Suppose for example that one wants to rewrite using a simple
hypothesis like %
\coqinline$H: forallx, x+-x = 0$ in a term like %
\coqinline$a+b+c+-(c+a)$. Since Coq standard \rewrite\ tactic matches
terms syntactically, this is not possible directly. Instead, one has
to reshape the goal using the commutativity and associativity lemmas:

\begin{twolistings}
\begin{coq}
\ttab rewrite (add_comm a b), <-(add_assoc b a c).
\ttab rewrite (add_comm c a), <-add_assoc.
\ttab rewrite H. 
\end{coq}&
\begin{coq}
(* |-~ ((a+b)+c)+-(c+a) = ... *)\ttab
(* |-~ (b+(a+c))+-(c+a) = ... *)
(* |-~ b+((a+c)+-(a+c)) = ... *)
(* |-~ b+0\ttab\ttab\ttab\tab= ... *)
\end{coq}
\end{twolistings}

\noindent
This is not satisfactory for several reasons. First, the proof script
is too verbose for such a simple reasoning step. Second, while reading
such a proof script is easy, writing it can be painful: there are
several sequences of rewrites yielding to the desired term, and
finding a reasonably short one is difficult. Third, we need to
copy-paste parts of the goal to select which occurrence to rewrite
using the associativity or commutativity lemmas; this is not a good
practice since the resulting script breaks when the goal is subject to
small modifications. (Note that one could also select occurrences by
their positions, but this is at least as difficult for the user which
then has to count the number of occurrences to skip, and even more
fragile since these numbers cannot be used to understand the proof
when the script breaks after some modification of the goal.)

In this paper, we propose a solution to this short-coming for the Coq
proof-assistant: we extend the usual rewriting tactic to automatically
exploit associativity and commutativity (AC), or just associativity
(A) of some operations.

\paragraph{Trusted unification vs untrusted matching.}

There are two main approaches to implementing rewriting modulo AC in a
proof-assistant.
First, one can extend the unification mechanism of the system to work
modulo AC~\cite{Pl72}. This option is quite powerful, since most
existing tactics would then work modulo AC. It however requires
non-trivial modifications of the kernel of the proof assistant (e.g.,
unification modulo AC does not always yield finite complete sets of
unifiers). As a consequence, this obfuscates the meta-theory: we need
a new proof of strong normalisation and we increase the trusted code
base.
On the contrary, we can restrict ourselves to pattern matching modulo
AC and use the core-system itself to validate all rewriting
steps~\cite{BoyerMoore81}. We chose this option.

\paragraph{Contributions, scope of the library.}

Besides the facts that such tools did not exist in Coq before and that
they apparently no longer exist in Isabelle/HOL
(see~\S\ref{ssec:related} for a more thorough discussion), the main
contributions of this work lie in the way standard algorithms and
ideas are combined together to get tactics that are efficient, easy to
use, and covering a large range of situations:
\begin{itemize}
\item We can have any number of associative and possibly commutative
  operations, each possibly having a neutral element. For instance, we
  can have the operations \code{min}, \code{max}, \code{+}, and
  \code{*} on natural numbers, where \code{max} and \code{+} share the
  neutral element \code{0}, \code{*} has neutral element \code{1}, and
  \code{min} has no neutral element.
\item We deal with arbitrary user-defined equivalence relations. This
  is important for rational numbers or propositions, for example,
  where addition and subtraction (resp.\ conjunction and disjunction)
  are not AC for Leibniz equality, but for rational equality,
  \code{Qeq} (resp.\ propositional equivalence, \code{iff}).
\item We handle ``uninterpreted'' function symbols: $n$-ary functions
  for which the only assumption is that they preserve the appropriate
  equivalence relation---they are sometimes called ``proper
  morphisms''.  For example, subtraction on rational numbers is a
  proper morphism for \code{Qeq}, while pointwise addition of
  numerators and denominators is not. (Note that any function is a
  proper morphism for Leibniz equality.)
\item The interface we provide is straightforward to use: it suffices
  to declare instances of the appropriate
  type-classes~\cite{SozeauOury08} for the operations of interest, and
  our tactics will exploit this information automatically.  Since the
  type-class implementation is first-class, this gives the ability to
  work with polymorphic operations in a transparent way (e.g.,
  concatenation of lists is declared as associative once and for all.)
\end{itemize}

\paragraph{Methodology.} 

Recalling the example from the beginning, an alternative to explicit
sequences of rewrites consists in making a transitivity step through a
term that matches the hypothesis' left-hand side syntactically:
\begin{twolistings}
\begin{coq}
transitivity (b+((a+c)+-(a+c))).
  ring. (* aac_reflexivity *)
rewrite H.
\end{coq}
&
\begin{coq}
(* |-`((a+b)+c)+-(c+a) = ... *)
(* \tab|-`((a+b)+c)+-(c+a) = b+((a+c)+-(a+c)) *)
(* |-`b+((a+c)+-(a+c)) = ... *)
(* |-`b+0 = ... *)
\end{coq}
\end{twolistings}

\noindent
Although the \code{ring} tactic~\cite{gregoire-mahboubi-05} solves the
first sub-goal here, this is not always the case (e.g., there are AC
operations that are not part of a ring structure). Therefore, we have
to build a new tactic for equality modulo A/AC: \tac.
Another drawback is that we also have to copy-paste and modify the
term manually, so that the script can break if the goal evolves. This
can be a good practice in some cases: the transitivity step can be
considered as a robust and readable documentation point; in other
situations we want this step to be inferred by the system, by pattern
matching modulo A/AC~\cite{hullot}.

All in all, we proceed as follows to define our \aacrewrite\ rewriting
tactic. Let $\eqac$ denote equality modulo A/AC; to rewrite using a
universally quantified hypothesis of the form $H\!: \forall \tilde{x},
p\tilde{x} = q\tilde{x}$ in a goal $G$, we take the following steps,
which correspond to building the proof-tree on the right-hand side:

\noindent
\begin{minipage}[t]{.55\linewidth}
\begin{enumerate}
\item \label{match-unif}choose a context $C$ and a substitution
  $\sigma$ such that $G \eqac C[p\sigma]$ (pattern matching modulo
  AC);
\item \label{make-trans}make a transitivity step through $C[p\sigma]$;
\item \label{close-dp} close this step using a dedicated
  decision procedure (\tac);
\item \label{rewrite}  use the standard \rewrite;
\item \label{continue} let the user continue the proof.
\end{enumerate}
\end{minipage}
\begin{minipage}[t]{.44\linewidth}
\begin{mathpar}
  \inferrule*[Right=\ref{make-trans}]{
    \inferrule*[Right=\ref{close-dp}]{ }
    {G \eqac C[p\sigma]} \\
    \inferrule*[Right=\ref{rewrite}]{
      H\and\inferrule*[Right=\ref{continue}]{\vdots}{C[q\sigma]}
    }{C[p\sigma]}
  }{G}
\end{mathpar}
\end{minipage}
\medskip

\noindent
For the sake of efficiency, we implement the first step as an OCaml
oracle, and we check the results of this (untrusted) matching function
in the third step, using the certified decision procedure \tac. To
implement this tactic, we use the standard methodology of
\emph{reflection}~\cite{BoyerMoore81,ACHA-reflection,gregoire-mahboubi-05}.
Concretely, this means that we implement the decision procedure as a
Coq function over ``reified'' terms, which we prove correct inside the
proof assistant.
This step was actually quite challenging: to our knowledge, \tac\ is
the first reflexive Coq tactic that handles uninterpreted function
symbols. In addition to the non-trivial reification process, a
particular difficulty comes from the (arbitrary) arity of these
symbols. To overcome this problem in an elegant way, our solution
relies on a dependently typed syntax for reified terms.

\paragraph{Outline.}

We sketch the user interface (\S\ref{sec:overall}) before describing
the decision procedure (\S\ref{sec:decision}) and the algorithm for
pattern matching modulo AC (\S\ref{sec:matching}). We detail our
handling of neutral elements and subterms separately
(\S\ref{sec:extend}). We conclude with related works and directions
for future work (\S\ref{sec:conclusion}).

\section{User interface, notation}
\label{sec:overall}

\paragraph{Declaring A/AC operations.}

We rely on type-classes~\cite{SozeauOury08} to declare the properties
of functions and A/AC binary operations. This allows the user to
extend both the decision procedure and the matching algorithm with new
A/AC operations or units in a very natural way. Moreover, this is the
basis of our reification mechanism (see~\S\ref{ssec:reification}).

\begin{figure}[t]
\begin{tabular}{c}\begin{coq}
Variables (X: Type) (R: relation X) (op: X -> X -> X).
Class Associative := law_assoc: forall x y z, R (op x (op y z)) (op (op x y) z).
Class Commutative := law_comm: forall x y, R (op x y) (op y x).
Class Unit (e: X) := { law_id_left: forall x, R (op e x) x; law_id_right: forall x, R (op x e) x }.  
\end{coq}\end{tabular}
\smallskip
\begin{twolistings}
\begin{coq}
Instance plus_A: Associative eq plus.
Instance plus_C: Commutative eq plus.
Instance plus_U: Unit eq plus O.
~
Instance app_A X: Associative eq (app X).
Instance app_U X: Unit eq (app X) (nil X).
\end{coq}
&
\begin{coq}
Instance and_A: Associative iff and.
Instance and_C: Commutative iff and.
Instance and_U: Unit iff and True.
Instance and_P: Proper (iff ==> iff ==> iff) and. 
Instance not_P: Proper (iff ==> iff) not. 
~
\end{coq}
\end{twolistings}
\caption{Classes for declaring properties of operations, example instances.}
\label{fig:classes}
\end{figure}

The classes corresponding to the various properties that can be
declared are given in Fig.~\ref{fig:classes}: being associative,
commutative, and having a neutral element.  Basically, a user only
needs to provide instances of these classes in order to use our
tactics in a setting with new A or AC operations.  These classes are
parameterised by a relation (\code{R}): one can use an arbitrary
equivalence relation.

Fig.~\ref{fig:classes} also contains examples of instances. 
Polymorphic values (\code{app}, \code{nil}) are declared in a
straightforward way. For propositional connectives (\code{and},
\code{not}), we also need to show that they preserve equivalence of
propositions (\code{iff}), since this is not Leibniz equality; we use
for that the standard \code{Proper} type-class---when the relation
\code{R} is Leibniz equality, these instances are inferred
automatically.
Of course, while we provide these instances, more can be defined by
the user.

\paragraph{Example usage.}

The main tactics we provide are \aacrewrite, to rewrite modulo
A/AC, and \aacreflexivity\, to decide an equality modulo A/AC.
Here is a simple example where we use both of them:
\begin{twolistings}
\begin{coq}
H1: forall x y z, x\capy \cup~x\capz = x\cap(y\cupz)
H2: forall x y, x\capx = x
a, b, c, d: set
=====================
(a\capc \cup~ b\capc\capd) \cap~c = (a \cup~d\capb) \cap~c
\end{coq}&
\begin{coq}
Proof.
 aac_rewrite H1; (* c`\cap`(a`\cup`b\capd)`\cap`c = ... *)
 aac_rewrite H2; (* c`\cap`(a`\cup`b\capd) \tab~`  = ... *)
 aac_reflexivity.
Qed.
\end{coq}
\end{twolistings}
\noindent
As expected, we provide variations to rewrite using the hypothesis
from right to left, or in the right-hand side of the goal.

\paragraph{Listing instances.} 

There might be several ways of rewriting a given equation: several
subterms may match, so that the user might need to select which
occurrences to rewrite. The situation can be even worse when rewriting
modulo AC: unlike with syntactical matching, there might be several
ways of instantiating the pattern so that it matches a given
occurrence. (E.g., matching the pattern $x+y+y$ at the root of the
term $a+a+b+b$ yields two substitutions: $\{x\mapsto a+a;y\mapsto b\}$
and the symmetrical one---assuming there is no neutral element.)  To
help the user, we provide an additional tactic, \aacinstances, to
display the possible occurrences together with the corresponding
instantiations. The user can then use the tactic \aacrewrite\ with the
appropriate options.

\paragraph{Notation and terminology.} 

We assume a signature $\Sigma$ and we let $f,g,h,\dots$ range over
function symbols, reserving letters $a,b,c,\dots$ for constants
(function symbols of arity $0$). We denote the set of \emph{terms} by
$T(\Sigma)$.
Given a set $V$ of variables, we let $x,y,z,\dots$ range over
(universally quantified) variables; a \emph{pattern} is a term with
variables, i.e., an element of $T(\Sigma + V)$.
A \emph{substitution} $(\sigma)$ is a partial function that maps
variables to terms, which we extend into a partial function from
patterns to terms, as expected.
Binary function symbols (written with an infix symbol, $\anysymb$) can
be associative (axiom $A$) and optionally commutative (axiom $C$);
these symbols may be equipped with a left and right unit $u$ (axiom
$U_{u,\anysymb}$):
\begin{align*}
A_\anysymb &: x \anysymb ( y \anysymb z ) \equiv (x\anysymb y)\anysymb z &
C_\anysymb &: x \anysymb y \equiv y \anysymb x &
U_{u,\anysymb} &: x \anysymb u \equiv  x  \wedge u \anysymb x  \equiv x %&
\end{align*}
We use $+_i$ (or $+$) for associative-commutative symbols (AC), and
$*_i$ (or $*$) for associative only symbols (A).
We denote by $\eqac$ the equational theory generated by these axioms
on $T(\Sigma)$. For instance, in a non-commutative semi-ring
$(+,*,0,1)$, $\eqac$ is generated by $A_{+},C_{+},A_{*}$ and
$U_{1,*},U_{0,+}$.

\section{Deciding equality modulo AC}
\label{sec:decision}

In this section, we describe the \tac\ tactic, which decides equality
modulo AC, is extensible through the definition of new type-class
instances, and deals with uninterpreted function symbols of arbitrary
arity.
For the sake of clarity, we defer the case where binary operations have
units to~\S\ref{ssec:units}.

\subsection{The algorithm and its proof}
\label{ssec:coqalgo}

\paragraph{A two-level approach.}  

We use the so called 2-level approach~\cite{barthe-barendregt-95}: we
define an inductive type \coqinline|T| for terms and a function
\coqinline|eval: T -> X| that maps reified terms to user-level terms
living in some type \coqinline|X| equipped with an equivalence
relation \code{R}, which we sometimes denote by $\equiv$. This allows
us to reason and compute on the syntactic representation of terms,
whatever the user-level model.

We follow the usual practice which consists in reducing equational
reasoning to the computation and comparison of normal forms: it then
suffices to prove that the normalisation function is correct to get a
sound decision procedure.
\begin{twolistings}
\begin{coq}
Definition norm: T -> T := ...
Lemma eval_norm: forall u, eval (norm u) \equiv eval u.
Theorem decide: forall u v, compare (norm u) (norm v) = Eq -> eval u \equiv eval v.
\end{coq}
\end{twolistings}
\noindent
This is what is called the \emph{autarkic way}: the verification is
performed inside the proof-assistant, using the conversion rule. To
prove \coqinline|eval u \equiv eval v|, it suffices to apply the
theorem \coqinline|decide| and to let the proof-assistant check by
computation that the premise holds.
The algorithm needs to meet two objectives. First, the normalisation
function (\code{norm}) must be efficient, and this dictates some
choices for the representation of terms. Second, the evaluation
function (\code{eval}) must be simple (in order to keep the proofs
tractable) and total: ill-formed terms shall be rejected
syntactically.

\paragraph{Packaging the reification environment.}

We need Coq types to package information about binary operations and
uninterpreted function symbols. They are given in Fig.~\ref{fig:envs},
where \coqinline{respectful} is the definition from Coq standard
library for declaring proper morphisms. We first define functions to
express the fact that $n$-ary functions are proper morphisms. A
``binary package'' contains a binary operation together with the
proofs that it is a proper morphism, associative, and possibly
commutative (we use the type-classes from Fig.~\ref{fig:classes}). An
``uninterpreted symbol package'' contains the arity of the symbol, the
corresponding function, and the proof that this is a proper morphism.

\begin{figure}[t]
\begin{coq}
(* type of n-ary homogeneous functions *)
Fixpoint type_of (X: Type) (n: nat): Type :=  
  match n with O =>~ X  | S n =>~ X -> type_of X n end.

(* relation to be preserved by n-ary functions *)
Fixpoint rel_of (X: Type) (R: relation X) (n: nat): relation (type_of X n) := 
  match n with O =>~ R  | S n =>~ respectful R (rel_of n) end.  
\end{coq} 

\begin{twolistings}
\begin{coq}
Module Bin.
 Record pack X R := {
  value: X -> X -> X;
  compat: Proper (R ==> R ==> R) value;
  assoc: Associative R value;
  comm: option (Commutative R value) }.        
End Bin.
\end{coq}
&
\begin{coq}
Module Sym.
 Record pack X R := {
  ar: nat;
  value: type_of X ar;
  compat: Proper (rel_of X R ar) value }.
~
End Sym.
\end{coq}
\end{twolistings}
\caption{Types for symbols.}
\label{fig:envs}
\end{figure}

The fact that symbols arity is stored in the package is crucial: by
doing so, we can use standard finite maps to store all function
symbols, irrespective of their arity. More precisely, we use two
environments, one for uninterpreted symbols and one for binary
operations; both of them are represented as non-dependent functions
from a set of indices to the corresponding package types:

\begin{tabular}{c}\begin{coq}
Variables (X: Type) (R: relation X).
Variable e_sym: idx -> Sym.pack X R.
Variable e_bin: idx -> Bin.pack X R.
\end{coq}\end{tabular}

\noindent
(The type \code{idx} is an alias for \code{positive}, the set of
binary positive numbers; this allows us to define the above functions
efficiently, using binary trees).

\paragraph{Syntax of reified terms.} 

We now turn to the concrete representation of terms.
The first difficulty is to choose an appropriate representation for AC
and A nodes, to avoid manipulating binary trees.  As it is usually
done, we flatten these binary nodes using variadic nodes. Since binary
operations do not necessarily come with a neutral element, we use
non-empty lists (resp. non-empty multi-sets) to reflect the fact that
A operations (resp. AC operations) must have at least one
argument.
(We could even require A/AC operations to have at least two
arguments but this would slightly obfuscate the code and prevent some
sharing for multi-sets.)
The second difficulty is to prevent ill-formed terms, like
\mbox{``\coqinline$log~1~2~3$''}, where a unary function is applied to
more than one argument. One could define a predicate stating that
terms are well-formed \cite{contejean-04}, and use this extra
hypothesis in later reasonings. We found it nicer to use dependent
types to enforce the constraint that symbols are applied to the right
number of arguments: it suffices to use vectors of arguments rather
than lists.
\begin{figure}[t]
\begin{twolistings}
\begin{coq}
(* non-empty lists/multisets *)
Inductive nelist A := 
| nil: A -> nelist A
| cons: A -> nelist A -> nelist A.

Definition nemset A := 
  nelist (A*positive).

(* reified terms *)
Inductive T: Type := 
| bin_ac: idx -> nemset T -> T
| bin_a : idx -> nelist T -> T
| sym: forall i, vect T (Sym.ar (e_sym i))  -> T.
\end{coq}  
&
\begin{coq}
Fixpoint eval (u: T): X := 
match u with
| bin_ac i l =>~let o:=Bin.value (e_bin i) in
   nefold_map o (fun(u,n)=>copy o n (eval u)) l
| bin_a i l =>~let o:=Bin.value (e_bin i) in
   nefold_map o eval l 
| sym i v =>~xeval v (Sym.value (e_sym i))
end
with xeval i (v: vect T i): Sym.type_of i->X := 
match v with
| vnil => (fun f =>~f)
| vcons u v => (fun f =>~xeval v (f (eval u)))
end.
\end{coq}
\end{twolistings}
\caption{Data-type for terms, and related evaluation function.}
\label{fig:terms}
\end{figure}
The resulting data-type for reified terms is given in
Fig.~\ref{fig:terms}; it depends on the environment for uninterpreted
symbols (\code{e_bin}). This definition allows for a simple
implementation of \code{eval}, given on the right-hand side. For
uninterpreted symbols, the trick consists in using an accumulator to
store the successive partial applications.

As expected, this syntax allows us to reify arbitrary user-level
terms. For instance, take \coqinline$(a*S(b+b))-b$. We first construct
the following environments where we store information about all atoms:
\begin{center}
  \begin{tabular}{cc|cc}
    \coqinlines|e_sym| &&& \coqinlines|e_bin| \\[.2em]
\begin{coq}
1 =>~{| ar := 1; value := S; compat := ...|}
2 =>~{| ar := 0; value := a; compat := ...|}
3 =>~{| ar := 0; value := b; compat := ...|}
_ =>~{| ar := 2; value := minus; compat := ...|}
\end{coq}
&&&
\begin{coq}
1 =>~{| value := plus; compat := ... ;
          assoc := _ ; comm := Some ... |}
_ =>~{| value := mult; compat := ... ;
          assoc := _ ; comm := None |}
\end{coq}
  \end{tabular}
\end{center}
These environment functions are total: they associate a semantic value
to indices that might be considered as ``out-of-the-bounds''. This
requires environments to contain at least one value, but this makes
the evaluation function total and easier to reason about: there is no
need to return an option or a default value in case undefined symbols
are encountered.
We can then build a reified term whose evaluation in the above
environments reduces to the starting user-level terms:
\begin{twolistings}\begin{coq}
Let t := sym 4 [|bin_a 2 [(sym 2 [||]); (sym 1 [|bin_ac 1 [(sym 3 [||],1);(sym 3 [||],1)|])]; sym 3 [||]|].
Goal eval e_sym e_bin t = (a*S(b+b))-b. reflexivity. Qed.
\end{coq}&\end{twolistings}

\noindent
Note that we cannot split the environment \coqinline|e_bin| into two
environments \coqinline|e_bin_a| and \coqinline|e_bin_ac|: since they
would contain at least one binary operation with the proof that it is
A or AC, it would not be possible to reify terms in a setting with
only A or only AC operations. Moreover, having a single environment
for all binary operations makes it easier to handle neutral elements
(see \S\ref{ssec:units}).

\paragraph{Normalisation of reified terms in Coq.}

Normal forms are computed as follows: terms are recursively flattened
under A/AC nodes and arguments of AC nodes are sorted.
We give excerpts of this Coq function below, focusing on AC nodes:
\coqinline|bin_ac'| is a smart constructor that prevents building
unary AC nodes, and \coqinline|norm_msets norm i| normalises and sorts a
multi-set, ensuring that none of its children starts with an AC node
with index~\code{i}.

\begin{coq}
Definition bin_ac' i (u: nemset T): T := match u with nil (u,1) =>~ u  | _ =>~ bin_ac i u end.
Definition extract_ac i (s: T): nemset T := 
  match s with bin_ac j m when i = j =>~ m | _ =>~ [s,1] end. 
Definition norm_msets norm i (u: nemset T): nemset T := 
  nefold_map merge_sort (fun (x,n) =>~ copy_mset n (extract_ac i (norm x))) u
...
Fixpoint norm (u: T): T := match u with
| bin_ac i l =>~ if is_commutative e_bin i then bin_ac' i (norm_msets norm i l)  else u
| bin_a i l =>~ bin_a' i (norm_lists norm i l)
| sym i l =>~ sym i (vector_map norm l)
end.
\end{coq}

Note that \code{norm} depends on the information contained in the
environments: the look-up 
\coqinline|is_commutative s_bin i| in the definition of
\coqinline|norm| is required to make sure that the binary operation
\coqinline|i| is actually commutative (remember that we need to store
A and AC symbols in the same environment, so that we might have AC
nodes whose corresponding operation is not commutative).  Similarly,
to handle neutral elements (\S\ref{ssec:units}), we will rely on the
environment to detect whether some value is a unit for a given binary
operation.

\paragraph{Correctness and completeness.}

We prove that the normalisation function is sound. This proof relies
on the above defensive test against ill-formed terms: since invalid AC
nodes are left intact, we do not need the missing commutativity
hypothesis when proving the correctness of \coqinline|norm|.
We did not prove completeness. First, this is not needed to get a
sound tactic. Second, this proof would be quite verbose (in
particular, it requires a formal definition of equality modulo AC on
reified terms). Third, we would not be able to completely prove the
completeness of the resulting tactic anyway, since one cannot reason
about the OCaml reification and normalisation functions in the
proof-assistant~\cite{gregoire-mahboubi-05,boutin-97}.

\subsection{Reification}
\label{ssec:reification} 

Following the reflexive approach to solve an equality modulo AC, it
suffices to apply the above theorem \coqinline|decide|
(\S\ref{ssec:coqalgo}) and to let Coq compute. To do so, we still need
to provide two environments \coqinline|e_bin| and \coqinline|e_sym|
and two terms \coqinline|u| and \coqinline|v|, whose evaluation is
convertible to the starting user-level terms.

\paragraph{Type-class based reification.}

We do not want to rely on annotations (like projections of
type-classes fields or canonical structures) to guess how to reify the
terms: this would force the users to use our definitions and notations
from the beginning.
Instead, we let the users work with their own definitions, and we
exploit type-classes to perform reification.
The idea is to query the type-class resolution mechanism to decide
whether a given subterm should be reified as an AC operation, an A
operation, or an uninterpreted function symbol.

The inference of binary A or AC operations takes place first, by
querying for instances of the classes \code{Associative} and
\code{Commutative} on all binary applications. The remaining
difficulty is to discriminate whether other applications should be
considered as a function symbol applied to several arguments, or as a
constant.
For instance, considering the application \coqinline|f (a+b)~(b+c)~c|,
it suffices to query for \coqinline|Proper| instances in the following
order:
\begin{center}
  \begin{tabular}{l@{\qquad}llr}
    1.&\coqinline|Proper (R ==>~R ==>~R ==>~R)|  & \coqinline|(f)| &~?\\
    2.&\coqinline|Proper (R ==>~R ==>~R)|  & \coqinline|(f (a+b))| & ?\\
    3.&\coqinline|Proper (R ==>~R)|  & \coqinline|(f (a+b) (b+c))| & ?\\
    4.&\coqinline|Proper (R)|  & \coqinline|(f (a+b) (b+c)~c)| & ?\\
  \end{tabular}
\end{center}
\noindent
The first query that succeeds tells which partial application is a
proper morphism, and with which arity. Since the relation \code{R} is
reflexive, and any element is proper for a reflexive relation, the
inference of constants---symbols of arity~0---is the catch-all case
of reification. We then proceed recursively on the remaining
arguments; in the example, if the second call is the first to succeed,
we do not try to reify the first argument (\code{a+b}): the partial
application \code{f(a+b)} is considered as an atom.

\paragraph{Reification language.} 

We use OCaml to perform this reification step. Using the meta-language
OCaml rather than the meta-language of tactics \ltac{} is a matter of
convenience: it allows us to use more efficient data-structures. For
instance, we use hash-tables to memoise queries to type-class
resolution, which would have been difficult to mimic in \ltac\ or
using canonical structures. 
The resulting code is non-trivial, but too technical to be presented
here. Most of the difficulties come from the fact that we reify
uninterpreted functions symbols using a dependently typed syntax, and
that our reification environments contain dependent records: producing
such Coq values from OCaml can be tricky.
Finally, using Coq's plug-in mechanism, we wrap up the previous ideas
in a tactic, \tac, which automates this process, and solves equations
modulo AC.

\paragraph{Efficiency.} 

The dependently typed representation of terms we chose in order to
simplify proofs does not preclude efficient computations.
The complexity of the procedure is dominated by the merging of sorted
multi-sets, which relies on a linear comparison function.
We did not put this decision procedure through an extensive testing;
however, we claim that it returns instantaneously in practice.
Moreover, the size of the generated proof is linear with respect to
the size of the starting terms. By contrast, using the tactic language
to build a proof out of associativity and commutativity lemmas would
usually yield a quadratic proof.

\section{Matching modulo AC}
\label{sec:matching}

Solving a matching problem modulo AC consists in, given a pattern $p$
and a term $t$, finding a substitution $\sigma$ such that $p\sigma
\eqac t$.  There are many known
algorithms~\cite{contejean-04,eker-02,hullot,nipkow-90}; we present
here a simple one.

\paragraph{Naive algorithm.}

Matching modulo AC can easily be implemented non-determi\-nistically.
For example, to match a sum $p_1+p_2$ against a term $t$, it suffices
to consider all possible decompositions of $t$ into a sum $t_1 +
t_2$. If matching $p_1$ against $t_1$ yields a solution (a
substitution), it can be used as an initial state to match $p_2$
against $t_2$, yielding a more precise solution, if any.
To match a variable $x$ against a term $t$, there are two cases
depending on whether or not the variable has already been assigned in
the current substitution.  If the variable has already been assigned
to a value $v$, we check that $v \eqac t$. If this is not the case,
the substitution must be discarded since $x$ must take incompatible
values. Otherwise, i.e.,\ if the variable is fresh, we add a mapping
from $x$ to $v$ to the substitution.
To match an uninterpreted node $f(\overline{q})$ against a term $t$,
it must be the case that $t$ is headed by the same symbol $f$, with
arguments $\overline{u}$; we just match $\overline{q}$ and
$\overline{u}$ pointwise.

\paragraph{Monadic implementation.}

We use a monad for non-deterministic and backtracking computations. 
Fig.~\ref{fig:search-monad-prim} presents the primitive functions
offered by this monad: \ocamlinline$>>=$ is a backtracking bind
operation, while \ocamlinline$>>|$ is non-deterministic choice.
\begin{figure}[t]
\begin{minipage}[t]{.4\linewidth}
\begin{tabular}{c}\begin{ocaml}
val (>>=): 'a m -> ('a -> 'b m) -> 'b m 
val (>>|): 'a m -> 'a m -> 'a m
val return: 'a -> 'a m   
val fail: unit -> 'a m
\end{ocaml}\end{tabular}
\caption{Search monad primitives.}
\label{fig:search-monad-prim}
\end{minipage}
\hfill
\begin{minipage}[t]{.5\linewidth}
\begin{tabular}{c}\begin{ocaml}
val split_ac: idx -> term -> (term * term) m
val split_a~: idx -> term -> (term * term) m
~
~
\end{ocaml}\end{tabular}
\caption{Search monad derived functions.} 
\label{fig:search-monad-derived}
\end{minipage}
\end{figure}
We have an OCaml type for terms similar to the inductive type we
defined for Coq reified terms: applications of A/AC symbols are
represented using their flattened normal forms. From the primitives of
the monad, we derive functions operating on terms
(Fig.~\ref{fig:search-monad-derived}): the function
\ocamlinline|split_ac i| implements the non-deterministic split of a
term $t$ into pairs $(t_1,t_2)$ such that $t \eqac t_1 +_i t_2$.
If the head-symbol of $t$ is $+_i$, then it suffices to split
syntactically the multi-set of arguments; otherwise, we return an
empty collection. The function \ocamlinline|split_a i| implements the
corresponding operation on associative only symbols.
\begin{figure}[t]
\begin{tabular}{c}\begin{ocaml}
mtch (p_1+_ip_2) t sigma = split_ac i t >>= (fun (t_1,t_2) -> mtch p_1 t_1 sigma >>= mtch p_2 t_2)
mtch (p_1*_ip_2) t sigma = split_a~ i t >>= (fun (t_1,t_2) -> mtch p_1 t_1 sigma >>= mtch p_2 t_2)
mtch (f(p_i)) (f(u_i)) sigma = fold_2 (fun acc p t -> acc >>= mtch p t) (return sigma) p_i u_i
mtch (var x) t sigma when Subst.find sigma x = None = return (Subst.add sigma x t)
mtch (var x) t sigma when Subst.find sigma x = Some v = if v \eqac t then return sigma else fail()
\end{ocaml}\end{tabular}
\caption{Backtracking pattern matching, using monads.}
\label{fig:deterministic-reduction}
\end{figure}
The matching algorithm proceeds by structural recursion on the
pattern, which yields the code presented in
Fig.~\ref{fig:deterministic-reduction} (using an informal ML-like
syntax). A nice property of this algorithm is that it does not produce
redundant solutions, so that we do not need to reduce the set of
solutions before proposing them to the user.

\paragraph{Correctness.} 

Following \cite{contejean-04}, we could have attempted to prove the
correctness of this matching algorithm. While this could be an
interesting formalisation work \emph{per se}, it is not necessary for
our purpose, and could even be considered an impediment. Indeed, we
implement the matching algorithm as an oracle, in an arbitrary
language.
Thus, we are given the choice to use a free range of optimisations,
and the ability to exploit all features of the implementation language.
In any case, the prophecies of this oracle, a set of solutions to the
matching problem, are verified by the reflexive decision procedure we
implemented in \S\ref{sec:decision}.

\section{Bridging the gaps}
\label{sec:extend}

Combining the decision procedure for equality modulo AC and the
algorithm for matching modulo AC, we get the tactic for rewriting
modulo AC. We now turn to lifting the simplifying assumptions we made
in the previous sections.

\subsection{Neutral elements}
\label{ssec:units}

Adding support for neutral elements (or ``units'') is of practical
importance:
\vspace{-.5em}
\begin{itemize}
\item to let \tac\ decide more equations (e.g.,
  \coqinline$max 0 (b*1)+a`=`a+b$);
\item to avoid requiring the user to
  normalise terms manually before performing rewriting steps (e.g., to
  rewrite using $\forall x,x\cup x=x$ in the term $a\cap
  b\cup\emptyset\cup b\cap a$);
\item to propose more solutions to
  pattern matching problems (consider rewriting $\forall x y,x\cdot
  y\cdot x^\perp=y$ in $a\cdot (b\cdot (a\cdot b)^\perp)$, where
  $\cdot$ is associative only with a neutral element: the variable $y$
  should be instantiated with the neutral element).
\end{itemize}

\paragraph{Extending the pattern matching algorithm.}

Matching modulo AC with units does not boil down to pattern matching
modulo AC against a normalised term: $a\cdot b\cdot (a\cdot b)^\perp$
is a normal form and the algorithm of
Fig.~\ref{fig:deterministic-reduction} would not give solutions with
the pattern $x\cdot y\cdot x^\perp$. The patch is however
straightforward: it suffices to let the non-deterministic splitting
functions (Fig.~\ref{fig:search-monad-derived}) use the neutral
element possibly associated with the given binary symbol. For
instance, calling \code{split_a} on the previous term would return the
four pairs
$\langle 1,a\cdot b\cdot (a\cdot b)^\perp\rangle$, %
$\langle a,b\cdot (a\cdot b)^\perp\rangle$, %
$\langle a\cdot b,(a\cdot b)^\perp\rangle$, and %
$\langle a\cdot b\cdot (a\cdot b)^\perp,1\rangle$, where $1$ is the
neutral element. %

\paragraph{Extending the syntax of reified terms.}

An obvious idea is to replace non-empty lists (resp. multi-sets) by
lists (resp.\ multi-sets) in the definition of
terms---Fig.~\ref{fig:terms}. This has two drawbacks. First, unless
the evaluation function (Fig.~\ref{fig:terms}) becomes a partial
function, every A/AC symbol must then be associated with a unit (which
precludes, e.g., \code{min} and \code{max} to be defined as AC
operations on relative numbers).  Second, two symbols cannot share a
common unit, like $0$ being the unit of both \code{max} and
\code{plus} on natural numbers: we would have to know at reification
time how to reify 0, is it an empty AC node for \code{max} or for
\code{plus}?
\begin{figure}[t]
\begin{twolistings}
\begin{coq}
Variable e_bin: idx -> Bin.pack X R      

Record binary_for (u: X) := {            
 bf_idx: idx;                                     
 bf_desc: Unit R (Bin.value (e_bin bf_idx)) u }.
\end{coq}
&
\begin{coq}
Record unit_pack := {
 u_value: X;
 u_desc: list (binary_for u_value) }.    

Variable e_unit: idx -> unit_pack.
\end{coq}
\end{twolistings}
\caption{Additional environment for terms with units.}
\label{fig:dec-syntax}
\end{figure}
Instead, we add an extra constructor for units to the
data-type of terms, and a third environment to store all units
together with their relationship to binary operations.
The actual definition of this third environment requires a more clever
crafting than the other ones. The starting point is that a unit is
nothing by itself, it is a unit for some binary operations.
Thus, the type of the environment for units has to depend on the
\code{e_bin} environment.
This type is given in Fig.~\ref{fig:dec-syntax}. The record
\coqinline|binary_for| stores a binary operation (pointed to by its
index \coqinline|bf_idx|) and a proof that the parameter \coqinline|u|
is a neutral element for this operation. Then, each unit is bundled
with a list of operations it is a unit for (\coqinline|unit_pack|):
like for the environment \code{e_sym}, these dependent records allow
us to use plain, non-dependent maps.
In the end, the syntax of reified terms depends only on the
environment for uninterpreted symbols (\code{e_sym}), to ensure that
arities are respected, while the environment for units (\code{e_unit})
depends on that for binary operations (\code{e_bin}).

\paragraph{Extending the decision tactic.}

Updating the Coq normalisation function to deal with units is fairly
simple but slightly verbose. Like we used the \code{e_bin} environment
to check that \code{bin_ac} nodes actually correspond to commutative
operations, we exploit the information contained in \code{e_unit} to
detect whether a unit is a neutral element for a given binary
operation.
On the contrary, the OCaml reification code, which is quite technical,
becomes even more complicated. Calling type-class resolution on all
constants of the goal to get the list of binary operations they are a
unit for would be too costly. Instead, we perform a first pass on the
goal, where we infer all A/AC operations and for each of these,
whether it has a neutral element. We construct the reified terms in a
second pass, using the previous information to distinguish units from
regular constants.

\subsection{Subterms}
\label{ssec:subterm} 

Another point of high practical importance is the ability to rewrite
in subterms rather than at the root.  Indeed, the algorithm of
Fig.~\ref{fig:deterministic-reduction} does not allow to match the
pattern $x+x$ against the terms $f(a+a)$ or $a+b+a$, where the
occurrence appears under some context. Technically, it suffices to
extend the (OCaml) pattern matching function and to write some
boilerplate to accommodate contexts; the (Coq) decision procedure is
not affected by this modification. Formally, subterm-matching a
pattern $p$ in a term $t$ results in a set of solutions which are
pairs $\langle C,\sigma\rangle$, where $C$ is a context and $\sigma$
is a substitution such that $C[p\sigma]\eqac t$.

\paragraph{Variable extensions.}

It is not sufficient to call the (root) matching function on all
syntactic subterms: the instance $a+a$ of the pattern $x+x$ is not a
syntactic subterm of $a+b+a$. The standard trick consists in enriching
the pattern using a \emph{variable
  extension}~\cite{PetersonS81:variable:extension,slind}, a variable
used to collect the trailing terms. In the previous case, we can
extend the pattern into $y+x+x$, where $y$ will be instantiated with
$b$.  It then suffices to explore syntactic subterms: when we try to
subterm-match $x+x$ against $(a+c)*(a+b+a)$, we extend the pattern
into $y+x+x$ and we call the matching algorithm
(Fig.~\ref{fig:deterministic-reduction}) on the whole term and the
subterms $a$, $b$, $c$, $a+c$ and $a+b+a$.  In this example, only the
last call succeeds.

\paragraph{The problem with subterms and units.}
\label{ssec:subtermunit}

However, this approach is not complete in the presence of
units. Suppose for instance that we try to match the pattern $x+x$
against $a*b$, where $*$ is associative only. If the variable $x$
can be instantiated with a neutral element $0$ for $+$, then the
variable extension trick gives four solutions:

\smallskip\hfill
 $a*b+[]$ \qquad\quad $(a+[])*b$ \qquad\quad $a*(b+[])$
\hfill\smallskip\\
(These are the returned contexts, in which $[]$ denotes the hole; the
substitution is always $\{x\mapsto 0\}$.) Unfortunately, if $*$ also
has a neutral element $1$, there are infinitely many other solutions:

\smallskip\hfill $a*b*(1+[])$ \qquad $a*b+0*(1+[])$ \qquad
$a*b+0*(1+0*(1+[]))$ \qquad \dots\quad
\hfill\smallskip\\
(Note that these solutions are distinct modulo AC, they collapse to
the same term only when we replace the hole with $0$.)  The latter
solutions only appear when the pattern can be instantiated to be equal
to a neutral element (modulo A/AC). We opted for a pragmatic solution
in this case: we reject these peculiar solutions, displaying a warning
message. The user can still instantiate the rewriting lemma
explicitly, or make the appropriate transitivity step using \tac.

\section{Conclusions}
\label{sec:conclusion}

The Coq library corresponding to the tools we presented is available
from~\cite{aac:web}. We do not use any axiom; the code consists of
about 1400 lines of Coq and 3600 lines of OCaml. We conclude with
related works and directions for future work.

\subsection{Related Works}
\label{ssec:related}

Boyer~and~Moore~\cite{BoyerMoore81} are precursors to our work in two
ways. First, their paper is the earliest reference to reflection we
are aware of, under the name ``Metafunctions''. Second, they use this
methodology to prove correct a simplification function for
cancellation modulo A.
By contrast, we proved correct a decision procedure for equality
modulo A/AC with units which can deal with arbitrary function symbols,
and we used it to devise a tactic for rewriting modulo A/AC.

\paragraph{Ring.}

While there is some similarity in their goals, our decision procedure
is incomparable with the Coq \ring\
tactic~\cite{gregoire-mahboubi-05}. On the one hand, \ring\ can make
use of distributivity and opposite laws to prove goals like
\coqinline$x^2-y^2 = (x-y)*(x+y)$, holding in any ring structure.
On the other hand, \tac\ can deal with an arbitrary number of AC or A
operations with their units, and more importantly, with uninterpreted
function symbols. For instance, it proves equations like
\coqinline$f(x\capy)`\cup`g(\emptyset\cup z)`=`g z`\cup`f(y\cap x)$, where
\code{f}, \code{g} are arbitrary functions on sets.
Like with \coqinline$ring$, we also have a tactic to normalise terms modulo
AC.

\paragraph{Rewriting modulo AC in HOL and Isabelle.}

Nipkow~\cite{nipkow-er} used the Isabelle system to implement
matching, unification and rewriting for various theories including AC.
He presents algorithms as proof rules, relying on the Isabelle
machinery and tactic language to build actual tools for equational
reasoning. While this approach leads to elegant and short
implementations, what is gained in conciseness and genericity is lost
in efficiency, and the algorithms need not terminate. 
The rewriting modulo AC tools he defines are geared toward automatic
term normalisation; by contrast, our approach focuses on providing the
user with tools to select and make one rewriting step efficiently.

Slind~\cite{slind} implemented an AC-unification algorithm and
incorporated it in the {\tt hol90} system, as an external and efficient
oracle. It is then used to build tactics for AC rewriting,
cancellation, and modus-ponens.
While these tools exploit pattern matching only, an application of
unification is in solving existential goals. 
Apart from some refinements like dealing with neutral elements and A
symbols, the most salient differences with our work are that we
use a reflexive decision procedure to check equality modulo A/AC
rather than a tactic implemented in the meta-language, and that we use
type-classes to infer and reify automatically the A/AC symbols and
their units.

Support for the former tool~\cite{nipkow-er} has been discontinued,
and it seems to be also the case for the latter~\cite{slind}. To our
knowledge, even though HOL-light and HOL provide some tactics to prove
that two terms are equal using associativity and commutativity of a
single given operation, tactics comparable to the ones we describe
here no longer exist in the Isabelle/HOL family of proof assistants.

\paragraph{Rewriting modulo AC in Coq.}

Contejean~\cite{contejean-04} implemented in Coq an algorithm for
matching modulo AC, which she proved sound and complete. The emphasis
is put on the proof of the matching algorithm, which corresponds to a
concrete implementation in the CiME system. Although decidability of
equality modulo AC is also derived, this development was not designed
to obtain the kind of tactics we propose here (in particular, we could
not reuse it to this end). Finally, symbols can be uninterpreted,
commutative, or associative and commutative, but neither associative
only symbols nor units are handled.

\smallskip

Gonthier et al.~\cite{gonthier:icfp11} have recently shown how to
exploit a feature of Coq's unification algorithm to provide ``less ad
hoc automation''. In particular, they automate reasoning modulo AC in
a particular scenario, by diverting the unification algorithm in a
complex but really neat way. Using their trick to provide the generic
tactics we discuss here might be possible, but it would be
difficult. Our reification process is much more complex: we have
uninterpreted function symbols, we do not know in advance which
operations are AC, and the handling of units requires a dependent
environment. Moreover, we would have to implement matching modulo AC
(which is not required in their example) using the same methodology;
doing it in a sufficiently efficient way seems really challenging.

\smallskip

Nguyen et al.~\cite{nguyen-kirchner-kirchner-02} used the external
rewriting tool ELAN to add support for rewriting modulo AC in
Coq. They perform term rewriting in the efficient ELAN environment,
and check the resulting traces in Coq. This allows one to obtain a
powerful normalisation tactic out of any set of rewriting rules which
is confluent and terminating modulo AC. Our objectives are slightly
different: we want to easily perform small rewriting steps in an
arbitrarily complex proof, rather than to decide a proposition by
computing and comparing normal forms.

The ELAN trace is replayed using elementary Coq tactics, and
equalities modulo AC are proved by applying the associativity and
commutativity lemmas in a clever way. On the contrary, we use the
high-level (but slightly inefficient) \rewrite\ tactic to perform the
rewriting step, and we rely on an efficient reflexive decision
procedure for proving equalities modulo AC. (Alvarado and Nguyen first
proposed a version where the rewriting trace was replayed using
reflection, but without support for modulo AC~\cite{alvarado}.)

From the user interface point of view, leaving out the fact that the
support for this tool has been discontinued, our work improves on
several points: thanks to the recent plug-in and type-class mechanisms
of Coq, it suffices for a user to declare instances of the appropriate
classes to get the ability to rewrite modulo AC. Even more
importantly, there is no need to declare explicitly all uninterpreted
function symbols, and we transparently support polymorphic operations
(like \code{List.app}) and arbitrary equivalence relations (like
\code{Qeq} on rational numbers, or \code{iff} on propositions).
It would therefore be interesting to revive this tool using the new
mechanisms available in Coq, to get a nicer and more powerful
interface.

\smallskip

Although this is not a general purpose interactive proof assistant,
the Maude system~\cite{maude}, which is based on equational and
rewriting logic, also provides an efficient algorithm for rewriting
modulo AC~\cite{eker-02}.
Like ELAN, Maude could be used as an oracle to replace our OCaml
matching algorithm. This would require some non-trivial interfacing
work, however. Moreover, it is unclear to us how to use these tools to
get all matching occurrences of a pattern in a given term.

\subsection{Directions for Future works.} 
\label{ssec:future}

\paragraph{Heterogeneous terms and operations.} 

Our decision procedure cannot deal with functions whose range and
domain are distinct sets. We could extend the tactic to deal with such
symbols, to make it possible to rewrite using equations like $\forall
u v, \|u+ v\| \leq \|u\| + \|v\|$, where $\|\cdot\|$ is a norm in a
vector space. This requires a more involved definition of reified
terms and environments to keep track of type information; the
corresponding reification process seems quite challenging.

We could also handle heterogeneous associative operations, like
multiplication of non-square matrices, or composition of morphisms in
a category. For example, matrix multiplication has type
\coqinline|forall`n`m`p, X n m -> X m p -> X n p| (\code{X n m} being
the type of matrices with size $n,m$). This would be helpful for
proofs in category theory. Again, the first difficulty is to adapt the
definition of reified terms, which would certainly require dependently
typed non-empty lists.

\paragraph{Other decidable theories.} 

While we focused on rewriting modulo AC, we could consider other
theories whose matching problem is decidable. Such theories include,
for example, the Abelian groups and the Boolean rings~\cite{boudet-89}
(the latter naturally appears in proofs of hardware circuits).

\paragraph{Integration with other tools.}

Recently, tactics have been designed to exploit external SAT/SMT
solvers inside Coq~\cite{SMTCoq}. These tactics rely on a reflexive
proof checker, used to certify the traces generated by the external
solver. However, in the SMT case, these traces do not contain proofs
for the steps related to the considered theories, so that one needs
dedicated Coq decision procedures to validate these steps. Currently,
mostly linear integer arithmetic is supported~\cite{SMTCoq}, using the
\code{lia} tactic~\cite{lia}; our tactic \aacreflexivity\ could be
plugged into this framework to add support for theories including
arbitrary A or AC symbols.

\section*{Acknowledgements}

We would like to thank Matthieu Sozeau for his precious help in
understanding Coq's internal API.

\bibliography{bib}

\end{document}